\documentclass[fleqn,12pt,twoside,eps]{article}
\usepackage[headings]{espcrc1}
\usepackage{epsfig}
\readRCS
$Id: espcrc1.tex,v 1.2 2004/02/24 11:22:11 spepping Exp $
\usepackage{graphicx}
\usepackage[figuresright]{rotating}

\newcommand{\AmS}{{\protect\the\textfont2
  A\kern-.1667em\lower.5ex\hbox{M}\kern-.125emS}}

\title{Two- and Three-Particle Jet Correlations from STAR}
\author{J. G. Ulery\address[Purdue]{Department of Physics, 
        Purdue University, West Lafayette, IN 47907, USA} (for the STAR\thanks{For full list of STAR authors and acknowledgments, see appendix `Collaborations' of this volume} Collaboration)%
        }
       
\runtitle{Two- and Three-Particle Jet Correlations from STAR}
\runauthor{J. G. Ulery}

\begin{document}

\maketitle

\begin{abstract}

Hard-soft charged hadron angular correlations are shown.  Away-side associated $\langle p_T\rangle$ is studied.  Angular correlations with trigger baryon and meson are compared.  Three-particle correlations are presented for the first time.  
\end{abstract}
%
%\section{INTRODUCTION}
\\

Relativistic heavy-ion collisions may create the quark-gluon plasma (QGP), a state of deconfined and thermalized quarks and gluons predicted by quantum chromodynamics (QCD) at high energy densities \cite{Karsch}.  To study the medium created in heavy-ion collisions we use jets.  Jets are used because they can be calculated by perturbative QCD and thus are well calibrated in the vacuum and because they can be modified by the medium and thus provide information about the medium \cite{XNWang}.

\section{TWO-PARTICLE AZIMUTHAL CORRELATIONS}

 \begin{figure}[htb]
\hfill
\begin{minipage}[t]{.5\textwidth}
    \begin{center}  
    	\vspace*{-0.001cm}
		\includegraphics[width=1.05\textwidth]{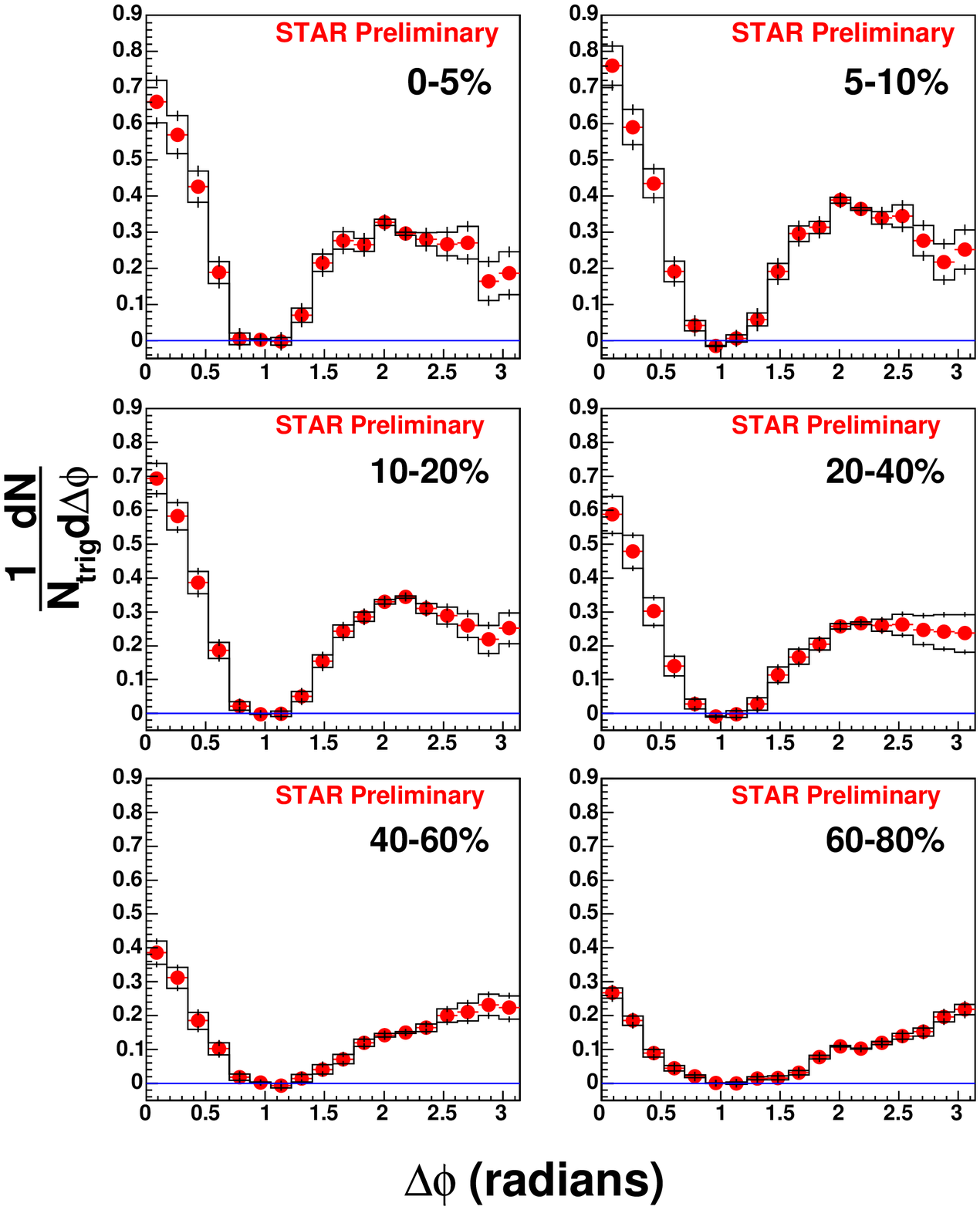}
		\vspace*{-1cm}
	\caption{$\Delta\phi$ correlations in Au+Au at different centralities with $2.5<p_T^{trig}<4.0$ GeV/c, $|\eta^{trig}|<0.7$ and $1.0<p_T<2.5$ GeV/c, $|\eta|<1.0$.  The histograms indicate systematic uncertainties.}
	\label{fig:phcomp}
	\end{center}
	\end{minipage}
	\hfill
\begin{minipage}[t]{.42\textwidth}
   \begin{center}  
   	\vspace*{-0.001cm}
   	\includegraphics[width=.7\textwidth]{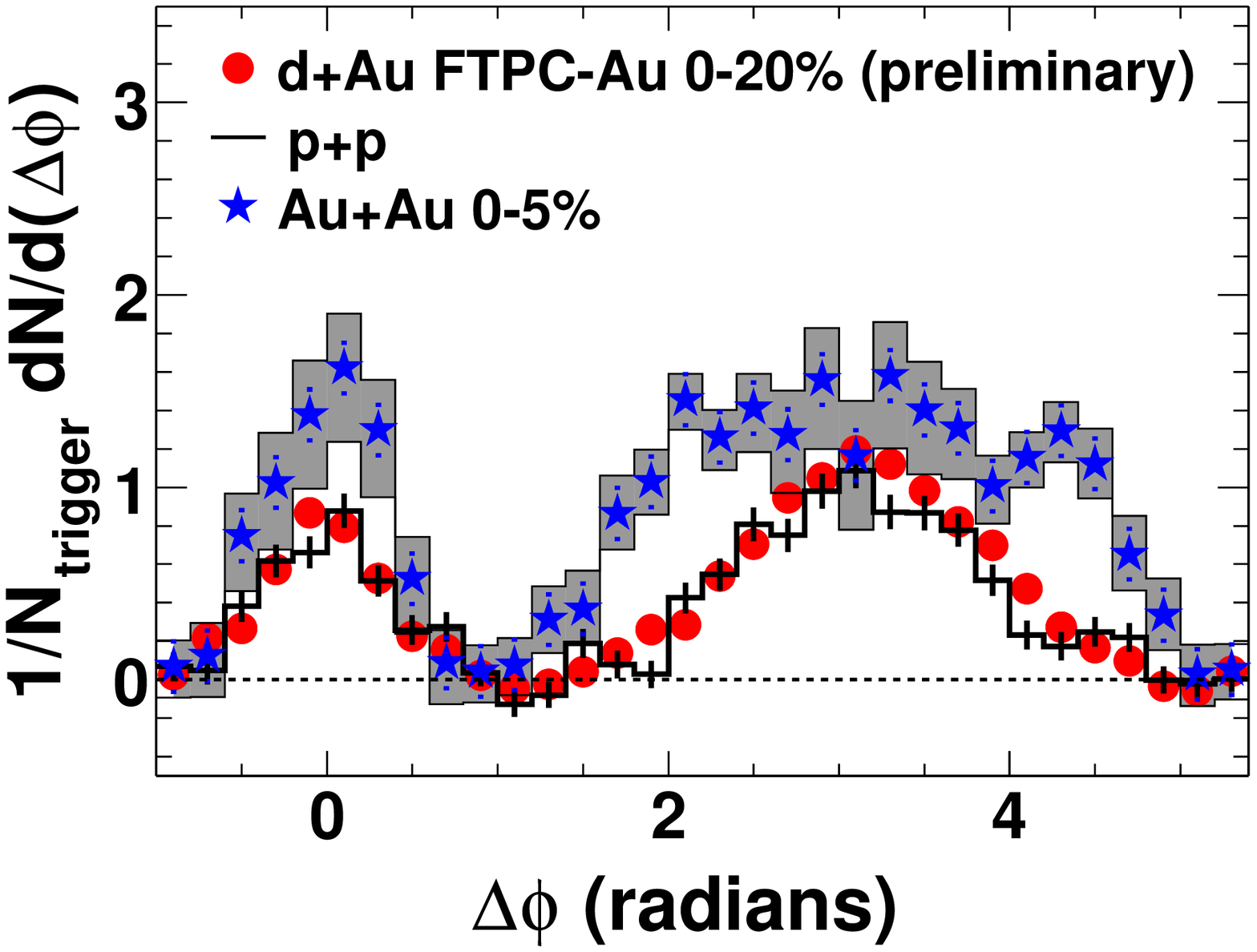}	\vspace*{-.45cm}
   	
   	\includegraphics[width=.7\textwidth]{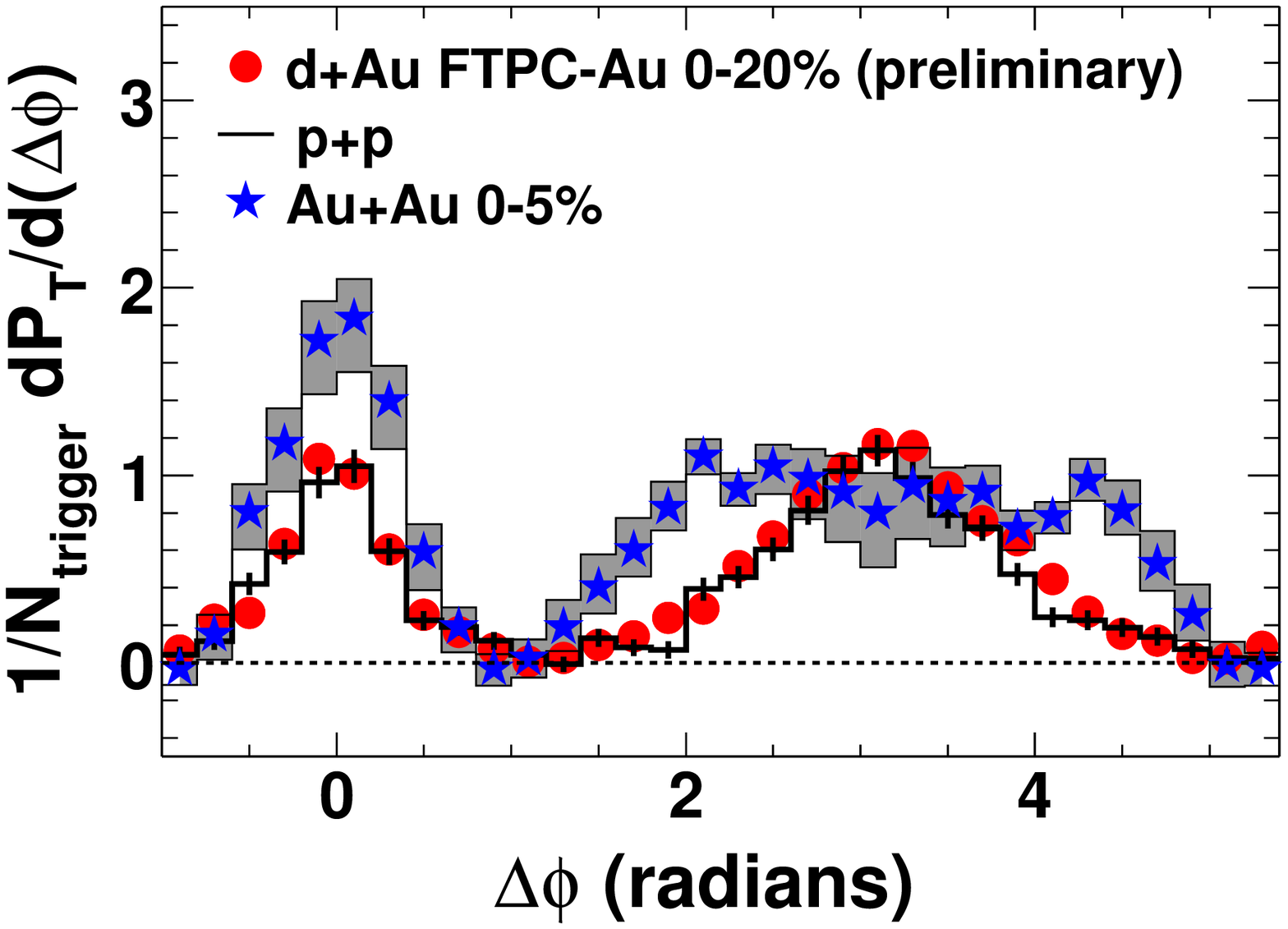}\vspace*{-.05cm}
   		\includegraphics[width=.8\textwidth]{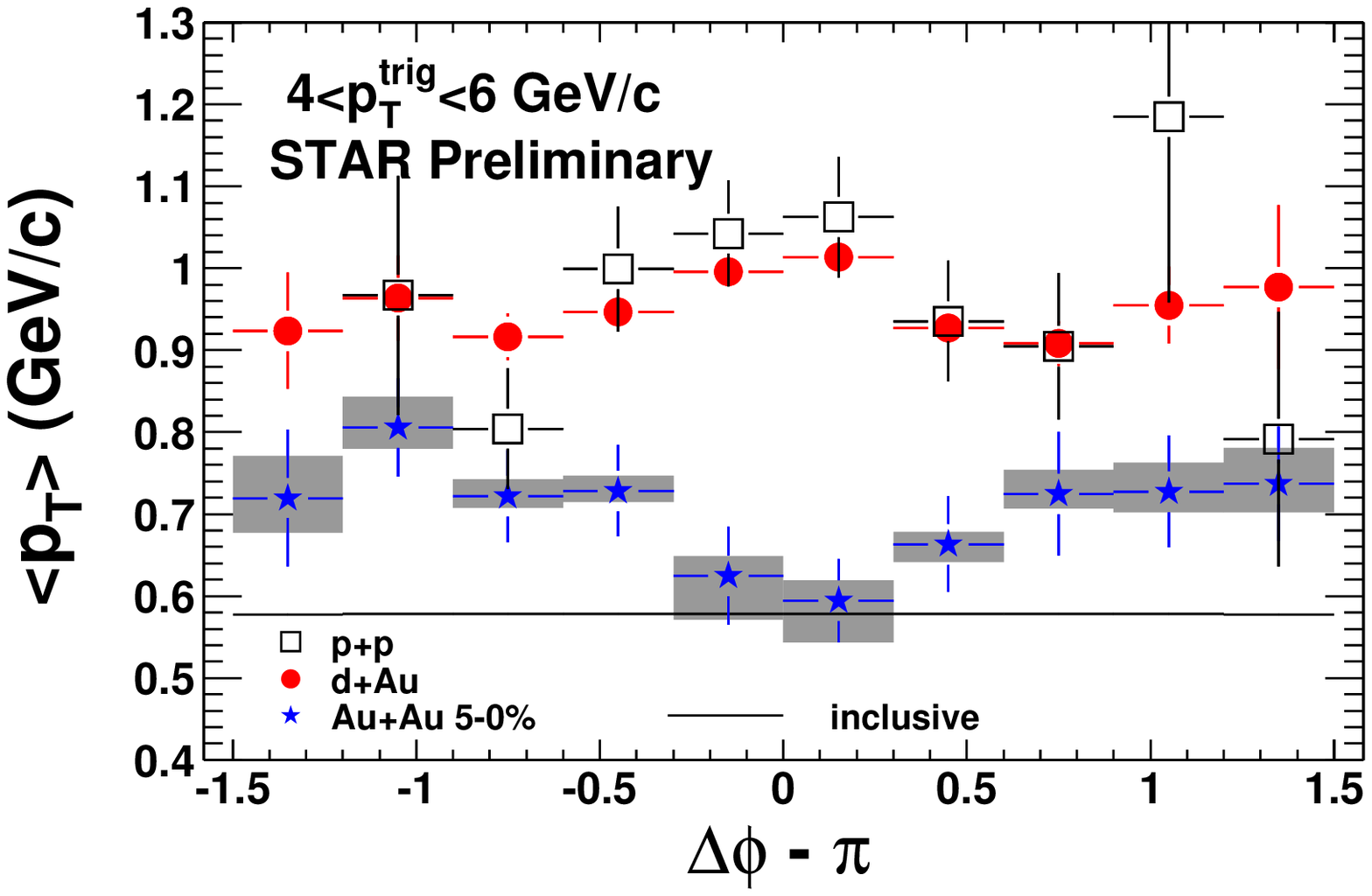}
   		\vspace*{-1cm}
	\caption{Top: $\Delta\phi$ correlations for $pp$, $20\%$ d+Au and $5\%$ Au+Au collisions with $4<p_T^{trig}<6$ GeV/c, $|\eta^{trig}|<0.7$ and $0.15<p_T<4.0$ GeV/c, $|\eta|<1.0$.  Middle:  Correlations weighted by $p_T$.  Bottom:  Away-side $\langle p_T \rangle$ versus $\Delta\phi$.  Shaded areas indicate systematic uncertainties.}
	\label{fig:pt}
	\end{center}
 \end{minipage}
\end{figure}

\begin{figure}[htb]
	\centering
		\includegraphics[width=.75\textwidth]{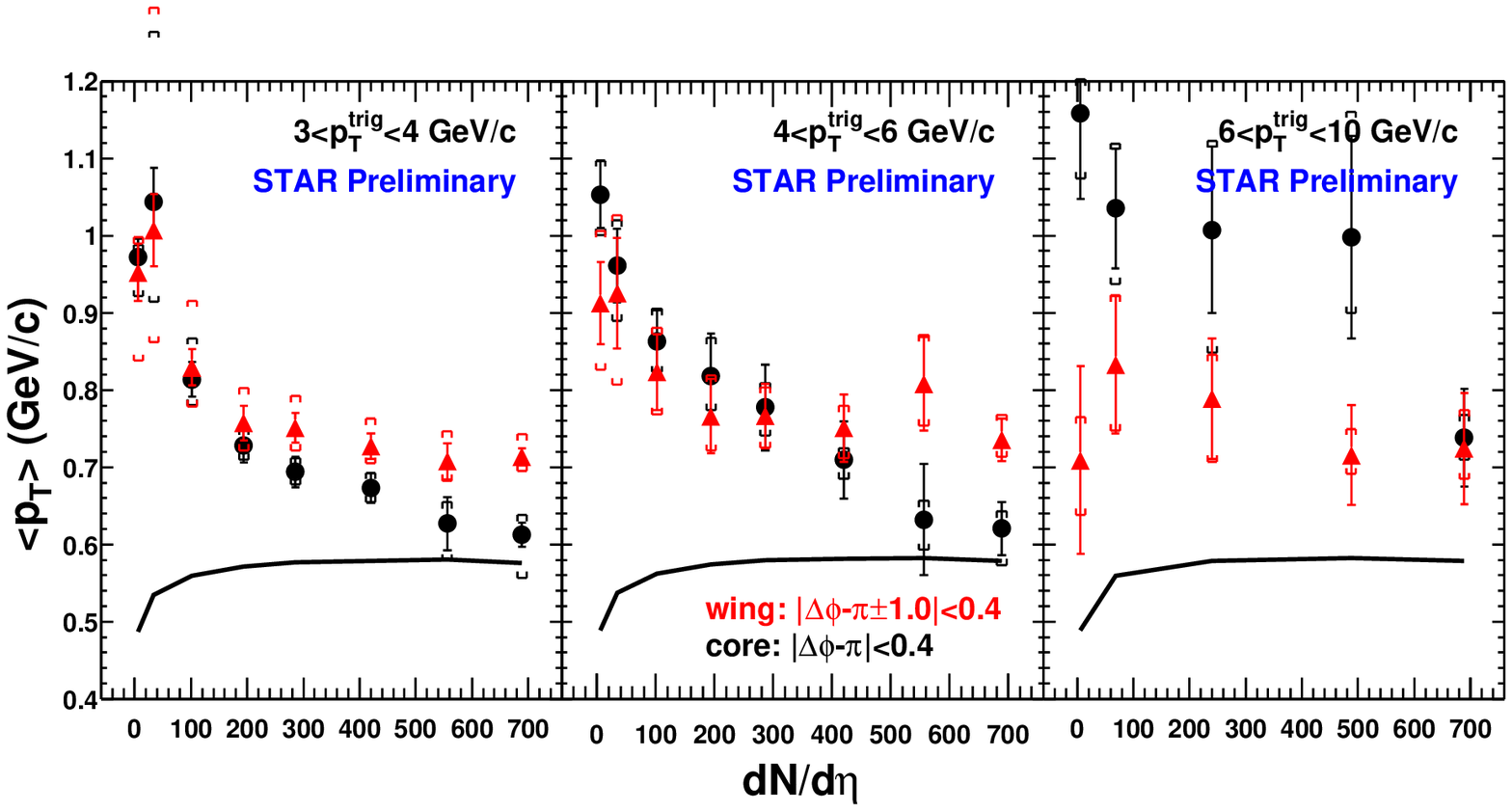}
	\vspace*{-1cm}
	\caption{(color online) Away-side associated particle $\langle p_T \rangle$ in the core, $|\Delta\phi-\pi|<0.4$, and the wing, $|\Delta\phi-(\pi\pm1)|<0.4$, as a function of centrality.  Caps indicated systematic uncertainties.  Curves are the inclusive $\langle p_T \rangle$.}
	\label{fig:MeanPtCent}
\end{figure}

Figure~\ref{fig:phcomp} shows jet-like angular correlations for six centrality bins in Au+Au collisions.  Trigger particles were selected to have transverse momentum $2.5<p_T^{trig}<4.0$ GeV/c and pseudo-rapidity $|\eta^{trig}|<0.7$.  Associated particles were selected to have $1.0<p_T<2.5$ GeV/c and $|\eta|<1.0$.  The away-side correlation evolves gradually from a peaked shape in peripheral to a flat (or even double-peaked) shape in central collisions, noting that the correlations are folded into $\Delta\phi=0-\pi$.

Figure~\ref{fig:pt} shows jet correlations of all charged hadrons, with and without $p_T$ weight, in Au+Au, d+Au and $pp$ collisions.  The d+Au and $pp$ results are similar, while in Au+Au correlations are enhanced on both the near and away side and the away-side peak is broadened. 
The ratio of the correlation functions yields the associated $\langle p_T \rangle$.  The $\langle p_T \rangle$ is seen to have a minimum at $\Delta\phi=\pi$ in central Au+Au while it peaks in $pp$ and d+Au.

Figure~\ref{fig:MeanPtCent} shows the centrality dependence of the away-side $\langle p_T \rangle$ in the core, $|\Delta\phi-\pi|<0.4$, and the wing, $|\Delta\phi-(\pi\pm1)|<0.4$, separately.  Both $\langle p_T \rangle$ decrease with centrality, consistent with \cite{JetSpectra}; however, the core decreases more rapidly, and appears more thermalized with the medium in central collisions.
%\section{IDENTIFIED TRIGGER PARTICLE CORRELATIONS}

\begin{figure}[htb]
	\centering
		\includegraphics[width=.98\textwidth]{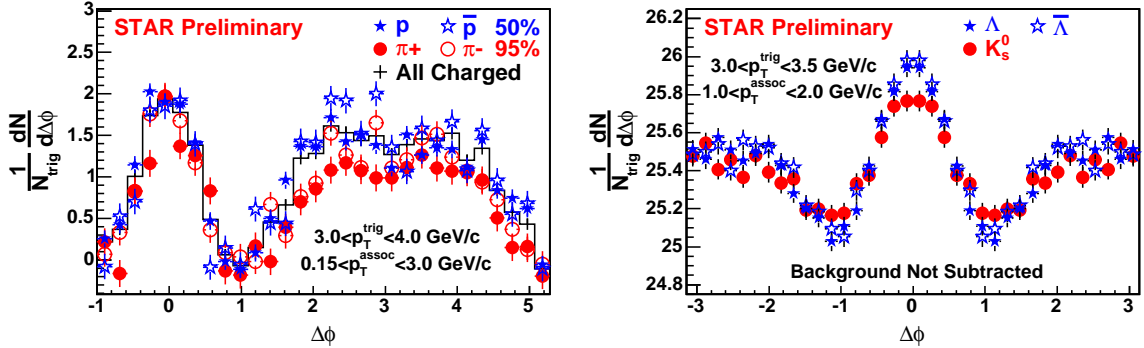}
			\vspace*{-1cm}
	\caption{(color online) Comparisons of $\Delta\phi$ correlations for identified trigger $\pi^+$, $\pi^-$, $p$, $\bar{p}$, $\Lambda$, $\bar{\Lambda}$ and $K^0_s$ in $5\%$ most central Au+Au collisions.  Error bars are statistical.}
	\label{fig:pid}
\end{figure}

Jet-like correlations with identified baryons and mesons may shed light on the baryon-meson puzzle \cite{BaryonMeson}.  Figure~\ref{fig:pid} shows jet-like correlations with intermediate $p_T$ trigger particles: $\pi^+$, $\pi^-$, $p$, and $\bar{p}$, identified by the relativistic rise of ionization energy loss, and topologically reconstructed $\Lambda$, $\bar{\Lambda}$ and $K^0_s$.  No significant difference is observed in the correlation between leading baryons and mesons, in contrast to expectations from coalescence/recombination models \cite{coal}.

\section{THREE-PARTICLE AZIMUTHAL CORRELATIONS}

\begin{figure}[htb]
\begin{center}
\hfill
	\begin{minipage}[t]{0.47\textwidth}
		\includegraphics[width=1.07\textwidth]{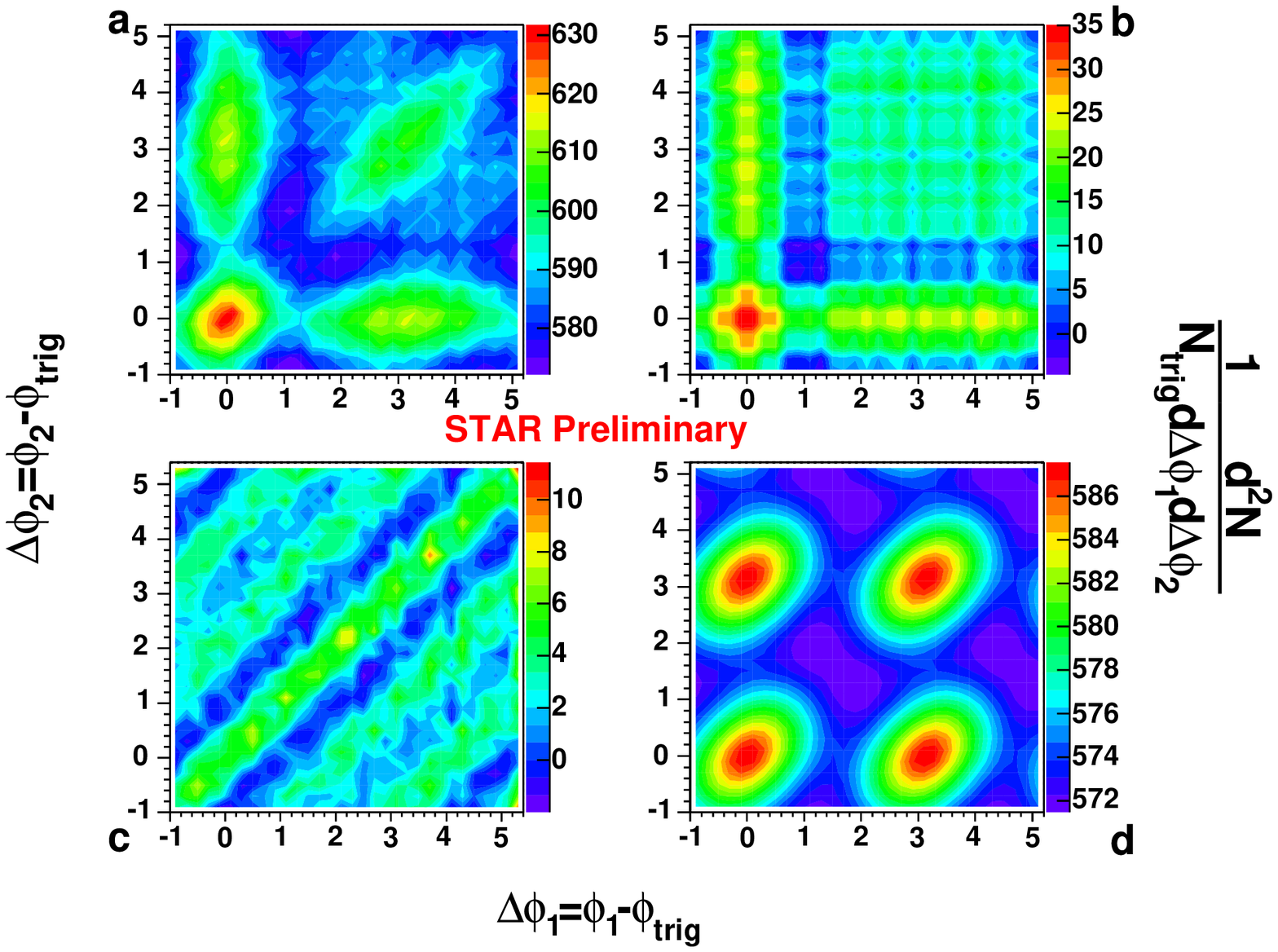}
		\vspace*{-1cm}
		\caption{(color online) Three-particle correlations in the $10\%$ most central Au+Au.  (a) Raw signal.  (b) Hard-soft background. (c) Soft-soft background.  (d) Random background.}
			\label{fig:3part3}
		\end{minipage}
		\hfill
		\begin{minipage}[t]{.47\textwidth}
		\includegraphics[width=1.07\textwidth]{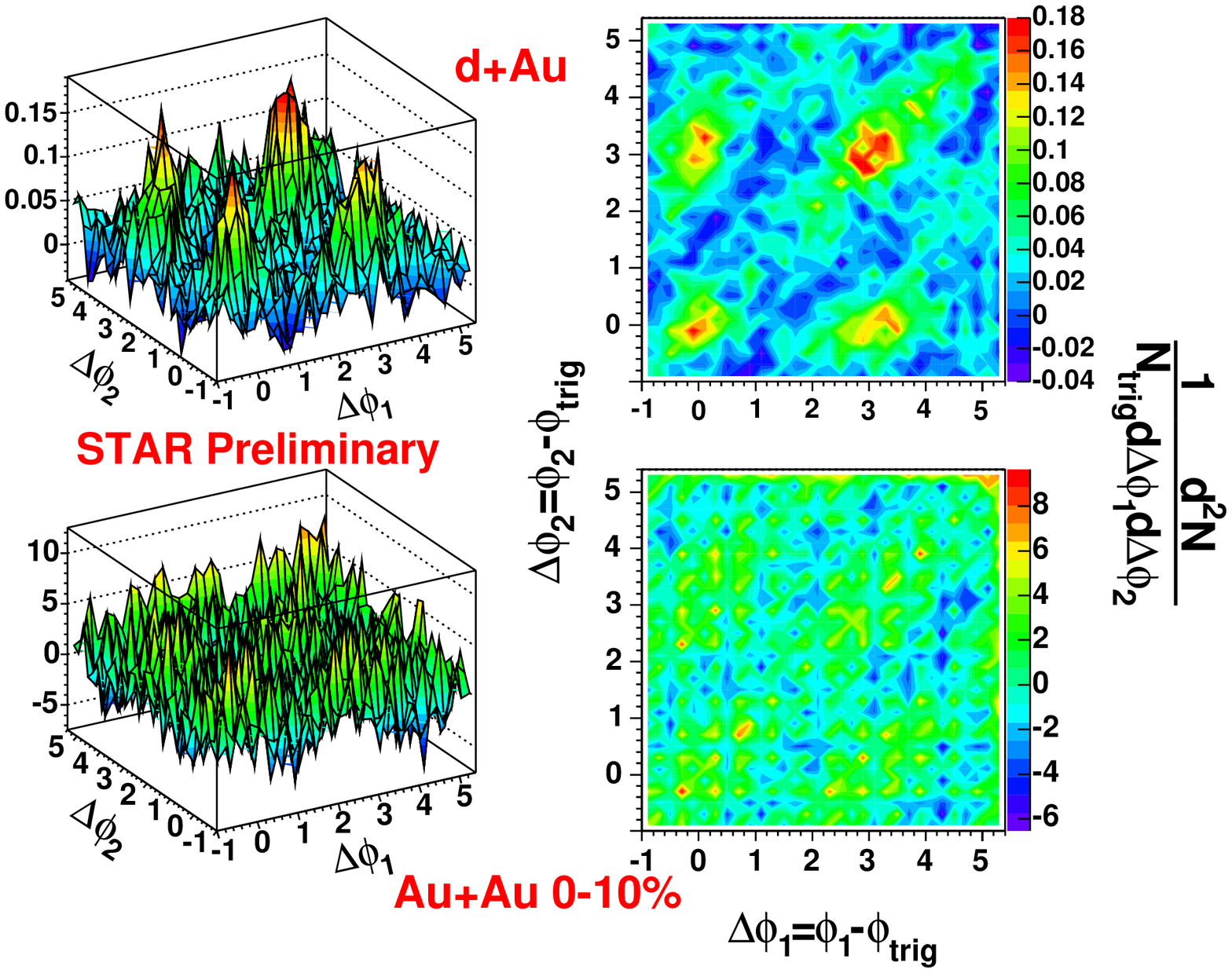}
			\vspace*{-1cm}
		\caption{(color online) Background subtracted 3-particle correlations.  Left and right panels are the same data plotted differently.  Top:  Minimum bias d+Au.  Bottom:  $10\%$ most central Au+Au.}
	\label{fig:Picture5}
	\end{minipage}
	\end{center}
\end{figure}

Recently it has been argued that the away-side flat (or even double-peaked) distributions are due to conical flow \cite{conical}.  This can be tested with 3-particle correlations.  Figure~\ref{fig:3part3}a shows the raw 3-particle correlation in $\Delta\phi_1$ = $\phi_1-\phi^{trig}$ and $\Delta\phi_2$ = $\phi_2-\phi^{trig}$ in $10\%$ most central Au+Au collisions.  A number of backgrounds need to be subtracted to isolate the 3-particle correlation due to jet-like sources.  One is a trigger-associated pair combined with a background particle; this is obtained by mixing 2-particle correlation with flow modulated background (Fig.~\ref{fig:3part3}b).  Another is a correlated pair in the underlying event that is unrelated to the trigger particle; this is obtained from soft particle correlations in events without a trigger particle (Fig.~\ref{fig:3part3}c).  The third is random background modulated by elliptic flow (taken to be the average of the 4-particle and reaction plane results):  $1+2v_2^{(1)}v_2^{trig}\cos(2\Delta\phi_1)+2v_2^{(2)}v_2^{trig}\cos(2\Delta\phi_2)+2v_2^{(1)}v_2^{(2)}\cos(2\Delta\phi_1-2\Delta\phi_2)$; normalization is fixed to the signal in strips of $0.8<|\Delta\phi_{1,2}|<1.2$ (Fig.~\ref{fig:3part3}d).

Figure~\ref{fig:Picture5} shows background subtracted 3-particle correlations between a trigger with $3<p_T^{trig}<4$ GeV/c, $|\eta^{trig}|<0.7$ and two associated particles with $1<p_T<2$ GeV/c, $|\eta|<1.0$.  Near-near $(0,0)$, away-away $(\pi,\pi)$, near-away $(0,\pi)$ and away-near $(\pi,0)$ peaks are evident.  An elongation is seen along the diagonal ($\Delta\phi_1=\Delta\phi_2$) in the d+Au away-side peak, perhaps due to $k_T$-broadening.  The away-side peak width along the diagonal is $0.89\pm0.11(stat.)$, while in the direction perpendicular to the diagonal it is $0.69\pm0.05$, consistent with the near-side peak width of $0.66\pm0.07$.

In Au+Au similar peaks are seen.  An elongation along the diagonal is also seen on the away side.  This could be a net-effect of $k_T$-broadening and jets deflected by medium radial flow.  The difference between the average signals in the central region, $|\Delta\phi_{1,2}-\pi|<0.4$, and the region of $|\Delta\phi_{1,2}-(\pi\pm1)|<0.4$ is $0.3\pm0.3(stat.)\pm0.4(sys.)$ radian$^{-2}$, indicating that the elongated structure is approximately flat.  Here the systematic uncertainties are from those in $v_2$ and background normalization.  In the presence of conical flow, one would expect peaks around $(\pi\pm1,\pi\mp1)$.  These peaks are {\em not} seen in the present data.  Difference between the average signals in the central region and the region of $|\Delta\phi_1-(\pi\pm1)|<0.4$ and $|\Delta\phi_2-(\pi\mp1)|<0.4$ is $2.6\pm0.3(stat.)\pm0.8(sys.)$ radian$^{-2}$.  

\section{CONCLUSIONS}

	Away-side jet-like dihadron correlations at intermediate $p_T$ show a gradual modification in correlation shape from peaked in peripheral to a broadened and flattened structure in central Au+Au collisions.
	The away-side $\langle p_T\rangle$ drops with centrality, more rapidly for associated particles close to $\pi$.   In central Au+Au collisions, the smallest value for the away-side $\langle p_T\rangle$ is found at $\Delta\phi=\pi$.  No significant difference is observed between correlations with baryon and meson trigger particles.  Three-particle correlations show an elongation along $\Delta\phi_1=\Delta\phi_2$ on the away side in both d+Au and Au+Au collisions.  No evidence of conical flow is observed in the present 3-particle correlation data.  These results test models describing high $p_T$ particle production.

\end{document}